\documentclass{aastex6}
\usepackage[utf8x]{inputenc}
\usepackage{graphicx}
\usepackage{multirow}
\usepackage{mathrsfs,amssymb}
\usepackage{amsmath}

\newcommand       \AU           {\,{\rm AU}}

\newcommand{\ee}[1]{\mbox{${} \times 10^{#1}$}}

\begin{document}
\pagestyle{plain}
\pagenumbering{arabic}

\title{Ionization-Driven Depletion and Redistribution of CO in
Protoplanetary Disks}

\author{Sarah E. Dodson-Robinson\altaffilmark{1}}
\author{Neal J. Evans II\altaffilmark{2},\altaffilmark{3}}
\author{Alyssa Ramos\altaffilmark{2}}
\author{Mo Yu\altaffilmark{2}}
\author{Karen Willacy\altaffilmark{4}}

\altaffiltext{1}{University of Delaware, Bartol Research
Institute/Department of Physics and Astronomy, 217 Sharp Lab, Newark, DE 19716}
\altaffiltext{2}{Astronomy Department, University of Texas, 1 University Station C1400, Austin, TX 78712, USA}
\altaffiltext{3}{Korea Astronomy and Space Science Institute, 776 Daedeokdae-ro, Yuseong-gu, Daejeon 34055, Korea}
\altaffiltext{4}{Mail Stop 169-506, Jet Propulsion Laboratory, California Institute of Technology, 4800 Oak Grove Drive, Pasadena, CA 91109}

\begin{abstract}



Based on the interstellar CO/H$_2$ ratio, carbon monoxide-based censuses
of protoplanetary disks in Lupus, $\sigma$~Orionis, and Chamaeleon I
found no disks more massive than the minimum-mass solar nebula, which is
inconsistent with the existence of exoplanets more massive than Jupiter.
Observations and models are converging on the idea that
ionization-driven chemistry depletes carbon monoxide in T-Tauri disks.
Yet the extent of CO depletion depends on the incident flux of ionizing
radiation, and some T-Tauri stars may have winds strong enough to shield
their disks from cosmic rays. There is also a range of X-ray
luminosities possible for a given stellar mass. Here we use a suite of
chemical models, each with a different incident X-ray or cosmic-ray
flux, to assess whether CO depletion is a typical outcome for T-Tauri
disks. We find that CO dissociation in the outer disk is a robust result
for realistic ionization rates, with abundance reductions between 70\%
and 99.99\% over 2~Myr of evolution. Furthermore, after the initial
dissociation epoch, the inner disk shows some recovery of the CO
abundance from CO$_2$ dissociation. In highly ionized disks, CO
recovery in the inner disk combined with depletion in the outer disk
creates a centrally peaked CO abundance distribution. The emitting area
in rare CO isotopologues may be an indirect ionization indicator: in a
cluster of disks with similar ages, those with the most compact CO
isotopologue emission see the highest ionization rates.


\end{abstract}


\maketitle

\section{Introduction}

The properties of protoplanetary disks set the conditions for planet
formation, and the workhorse molecule for determining disk masses and
turbulent speeds has been CO and its rarer isotopologues. Observations
are sometimes interpreted by assuming that all carbon is in gas-phase CO
inside a freeze-out radius \citep[e.g.][]{williams14, ansdell16,
ansdell17}. However, the chemical models of
\citet{aikawa97,bruderer12,favre13,yu16} predict that CO abundances are
much lower than the interstellar value even inside the freeze-out
radius, an effect we call chemical depletion. Along with other factors
that affect the emission from commonly observed CO transitions, this
chemical depletion can lead to underestimation of disk masses by large
factors \citep{miotello16,miotello17,molyarova17,yu17a}. The chemical
depletion also affects tracers of disk turbulence like the
peak-to-trough ratio; not accounting for chemical depletion can cause
disk turbulence to be underestimated \citep{yu17b}. Turbulent
diffusion can also deplete CO from the warm molecular layer, as CO gas
from downward-sinking eddies freezes onto the grains, which grow too
large to diffuse back up into the warm molecular layer and release the
CO \citep{xu17}. Significant CO depletion has been seen in other models
\citep{furuya14,reboussin15,drozdovskaya16,eistrup16,eistrup18,schwarz18};
though differing in details, the models agree that CO is likely to be
depleted inside the freeze-out radius.

The conclusion that CO is chemically depleted depends on the disk's
chemical composition evolving on a timescale comparable to disk
lifetime. The CO chemial evolution is driven by ionization from cosmic
rays, X-rays, and radionuclide decay. Yet the cosmic ray ionization rate
inferred from primitive solar nebula abundances \citep{umebayashi09} may
not represent all protoplanetary disks: \citet{cleeves13a} suggest that
magnetic fields in stellar winds can create a ``T-Tauriosphere'',
screening cosmic rays. Also, young stars have a wide range of X-ray
fluxes \citep{garmire00}, and X-ray luminosity may vary in time
\citep{robrade06,robrade07}. Our goal is to test how a disk's CO
abundance distribution depends on the dominant ionization mechanism and
the ionization rate.

Here we present a suite of models of chemically evolving protostellar
disks with different ionization rates. All other disk properties are
kept constant. In \S \ref{sec:models}, we describe the ionization rates
$\zeta(R,z)$ used in our calculations. In \S \ref{sec:abundances}, we
discuss the CO abundance calculated for each ionization profile.  In \S
\ref{sec:distribution}, we demonstrate that increasing CO
abundance in the inner disk combined with CO depletion beyond
20~au, but inside the CO ice line occurs for all of our models, which
have ionization rates consistent with X-ray observations and inferred
cosmic-ray fluxes for the solar nebula and nearby planet-forming disks.
We present our conclusions in \S \ref{sec:conclusions}.


\section{Disk Models and Ionization Rates}
\label{sec:models}

Our model disk has mass $0.015 M_{\odot}$ contained within a 70~AU
radius. Disk evolution begins when the central star, which has mass $1
M_{\odot}$, is 0.1~Myr old, roughly the beginning of the T-Tauri phase
\citep[e.g.][]{kristensen18}, and the star evolves along the Hayashi
track throughout the 3~Myr of disk evolution according to the models of
\citet{dantona94}. The fiducial model rates for ionization are fully
described in \citet{yu16}. Here we summarize them. Ionization by
ultraviolet radiation is negligible except for the very surface layer of
the disk. X-rays are able to reach most of the disk, producing an
ionization rate around ${10}^{-17}$~s$^{-1}$ in the disk interior,
except for the dense inner $10\AU$ near the disk midplane. If magnetic
shielding is neglected, cosmic rays provide an ionization rate of
${10}^{-17}$~s$^{-1}$ in nearly all of the disk, with only minor
attenuation in the inner midplane. We experiment with decreases of
factors of 10 to 100 in both X-rays and cosmic rays.  We also consider
{\bf increases} of a factor of 10 in each. Decreases in the total
ionization rate by more than a factor of 100 are not realistic because
the decay of short-lived radionuclides (SLRs) such as $^{26}$Al can
provide an ionization rate on the order of ${10}^{-19}$~s$^{-1}$ to
${10}^{-18}$~s$^{-1}$ \citep{cleeves13b}. Roughly speaking, X-rays
dominate the ionization rate where the vertical column density is less
than a few g~cm$^{-2}$, while cosmic-ray ionization dominates the disk
interior.

\begin{figure}[ht]
\centering
\includegraphics[width=0.9\textwidth]{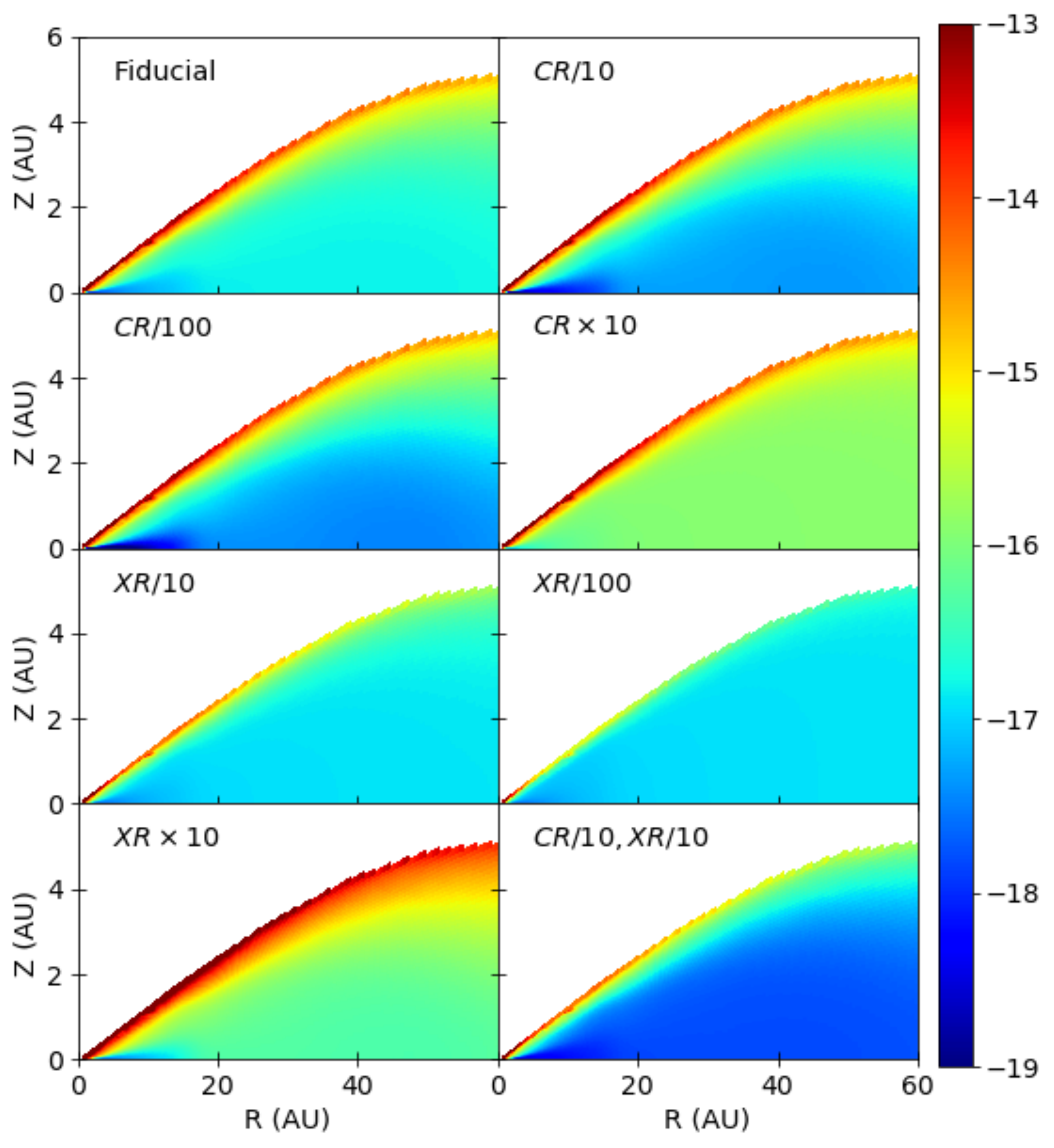}
\caption{Total ionization rate from the combination of X-rays
and cosmic rays in each model. See Table \ref{ratestab} for an
explanation of the model names.}
\label{fig: Ionization}
\end{figure}

\begin{figure}[ht]
\centering
\includegraphics[width=0.5\textwidth]{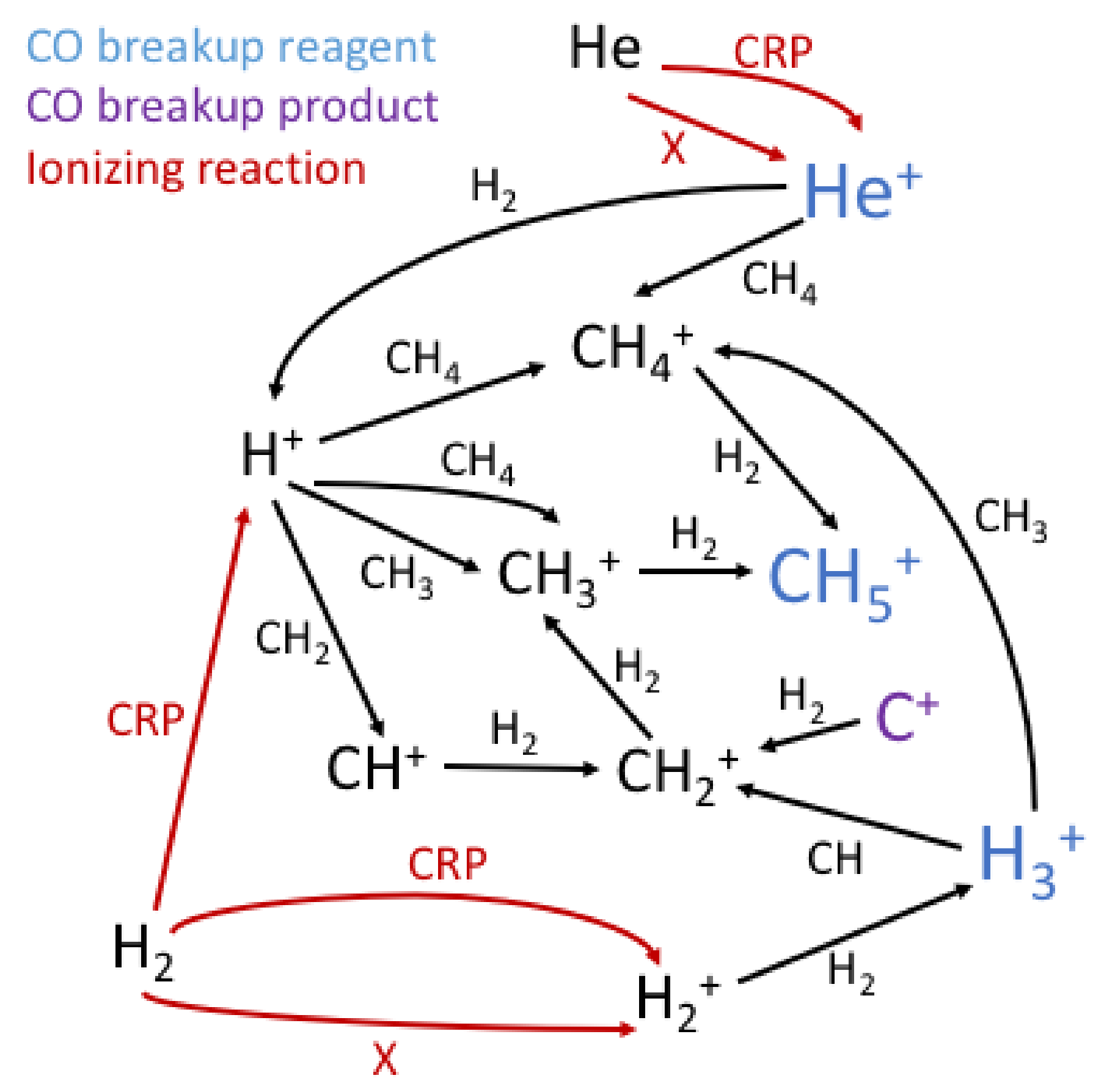}
\caption{Network of ion exchange reactions that creates He$^+$,
CH$_5^+$, and H$_3^+$, the molecules primarily responsible for
CO destruction (Equations \ref{eq:coheplus}--\ref{eq:coch5plus}).
Initial cosmic ray- or X-ray-driven ionizations are shown in
red, while the CO-destroying molecules are shown in blue. C$^+$,
which is produced by the breakup of CO in reaction
\ref{eq:coheplus}, can further drive CO destruction by reacting
with H$_2$ to form CH$_2^+$.}
\label{fig: Network}
\end{figure}

Figure \ref{fig: Ionization} shows the total ionization rate
$\zeta(R,z)$, including both cosmic rays and X-rays, as a function of
disk location. We do not consider time-varying ionization:
$\zeta(R,z)$ stays constant in time, consistent with a star that lacks
accretion bursts and is not near evolving massive stars. Future work
should at least explore time-varying X-ray flux, as was observed by
\citet{robrade06} and \citet{robrade07}.  A numerical summary of the
models at a single location in the disk, $R = 38$~au and $z = 0$, is
given in Table \ref{ratestab}.  The first column lists the name for the
model, and the rest of the columns show the rates for ionizing reactions
\ref{eq:h2crp1}--\ref{eq:hex} (listed below) and the fractional electron
abundance $f(e) = N_{\rm e^-} / \left(N_{\rm H} + 2 N_{\rm H_2} \right)$
at $t = 0$. These rates and free electron abundances allow simple
comparisons of the effects of the different models. The ionizing
reactions that initiate the chemical reaction network leading to CO
chemical depletion are:
\begin{equation}
\rm{CRP} + H_2 \rightarrow\ H^+ + H + e^-
\label{eq:h2crp1}
\end{equation}
\begin{equation}
\rm{CRP} + H_2 \rightarrow\ H_2^+ + e^-
\label{eq:h2crp2}
\end{equation}
\begin{equation}
\rm{X} + H_2 \rightarrow\ H_2^+ + e^-
\label{eq:h2x}
\end{equation}
\begin{equation}
\rm{CRP} + He \rightarrow\ He^+ + e^-
\label{eq:hecrp}
\end{equation}
\begin{equation}
\rm{X} + He \rightarrow\ He^+ + e^-,
\label{eq:hex}
\end{equation}
where CRP is a cosmic-ray particle and X is an X-ray.
Figure \ref{fig: Network} shows the reactions in the ion chemistry
network that affect the CO abundance. Input of ionizing particles is
shown in red. The three ions shown in blue destroy the CO
molecule:
\begin{equation}
{\rm He^+ + CO \rightarrow\ O + C^+ + He}
\label{eq:coheplus}
\end{equation}
\begin{equation}
{\rm H_3^+ + CO \rightarrow\ HCO^+ + H_2}
\label{eq:coh3plus}
\end{equation}
\begin{equation}
{\rm CH_5^+ + CO \rightarrow\ HCO^+ + CH_4}
\label{eq:coch5plus}
\end{equation}
While He$^+$ is produced directly from cosmic-ray or X-ray ionization
of helium, CH$_5^+$ and H$_3^+$ come from charge exchanges
initiated by ionization products H$^+$ and H$_2^+$.
Although the HCO$^+$ produced by reactions \ref{eq:coh3plus} and
\ref{eq:coch5plus} can be recycled to re-form CO (e.g.\ ${\rm HCO^+ +
e^- \rightarrow CO + H}$ or ${\rm HCO^+ + C_2H_2 \rightarrow CO +
C_2H_3^+}$), we find that CO destruction dominates throughout most of
the disk. C$^+$, a product of reaction \ref{eq:coheplus} (shown in
purple in Figure \ref{fig: Network}), can feed back into the CO
destruction network by reacting with H$_2$. Reaction products
that do not participate directly in ionization-driven CO
destruction are not shown in Figure \ref{fig: Network}.

\begin{table}[h]
\caption{Rates and Electron Abundances} \label{ratestab} 
\vspace {3mm}
\begin{tabular}{l r r r r r r}
\tableline
\tableline
Model\footnotemark[1] & Rate \ref{eq:h2crp1} & Rate \ref{eq:h2crp2} & Rate
\ref{eq:h2x} & Rate \ref{eq:hecrp} & Rate \ref{eq:hex} & f(e) \cr
  & s$^{-1}$    & s$^{-1}$  & s$^{-1}$  &  s$^{-1}$ & s$^{-1}$ & \cr 
\tableline
Fiducial   & 1.37\ee{-09} & 7.48\ee{-08} & 2.79\ee{-08} &
1.14\ee{-08} & 7.80\ee{-09} & 9.64\ee{-12} \cr
CR/10      & 1.37\ee{-10} & 7.48\ee{-09} & 2.79\ee{-08} &
1.14\ee{-09} & 7.80\ee{-09} & 4.42\ee{-12} \cr
CR/100     & 1.37\ee{-11} & 7.48\ee{-10} & 2.79\ee{-08} &
1.14\ee{-10} & 7.80\ee{-09} & 3.26\ee{-12} \cr
CR$\times$10 & 1.37\ee{-08} & 7.48\ee{-07} & 2.79\ee{-08} &
1.14\ee{-07} & 7.80\ee{-09} & 2.27\ee{-11} \cr
XR/10      & 1.37\ee{-09} & 7.48\ee{-08} & 2.79\ee{-09} &
1.14\ee{-08} & 7.80\ee{-10} & 9.36\ee{-12} \cr
XR/100     & 1.37\ee{-09} & 7.48\ee{-08} & 2.79\ee{-10} &
1.14\ee{-08} & 7.80\ee{-11} & 9.32\ee{-12} \cr
XR$\times$10 & 1.37\ee{-09} & 7.48\ee{-08} & 2.79\ee{-07} &
1.14\ee{-08} & 7.80\ee{-08} & 1.20\ee{-11} \cr
CR/10,XR/10  & 1.37\ee{-10} & 7.48\ee{-09} & 2.79\ee{-09} &
1.14\ee{-09} & 7.80\ee{-10} & 3.45\ee{-12} \cr

\tableline
\footnotetext[1]{Model names syntax: CR/10 has 1/10$^{th}$ of the
cosmic ray-driven ionization rate and the same X-ray-driven
ionization rate as the fiducial model; XR$\times$10 has the same
cosmic ray-driven ionization rate and 10 times the X-ray-driven
ionization rate as the fiducial model; etc.}

\end{tabular}
\end{table}

\section{CO abundances}
\label{sec:abundances}

\begin{figure}[ht]
\centering
\includegraphics[width=0.9\textwidth]{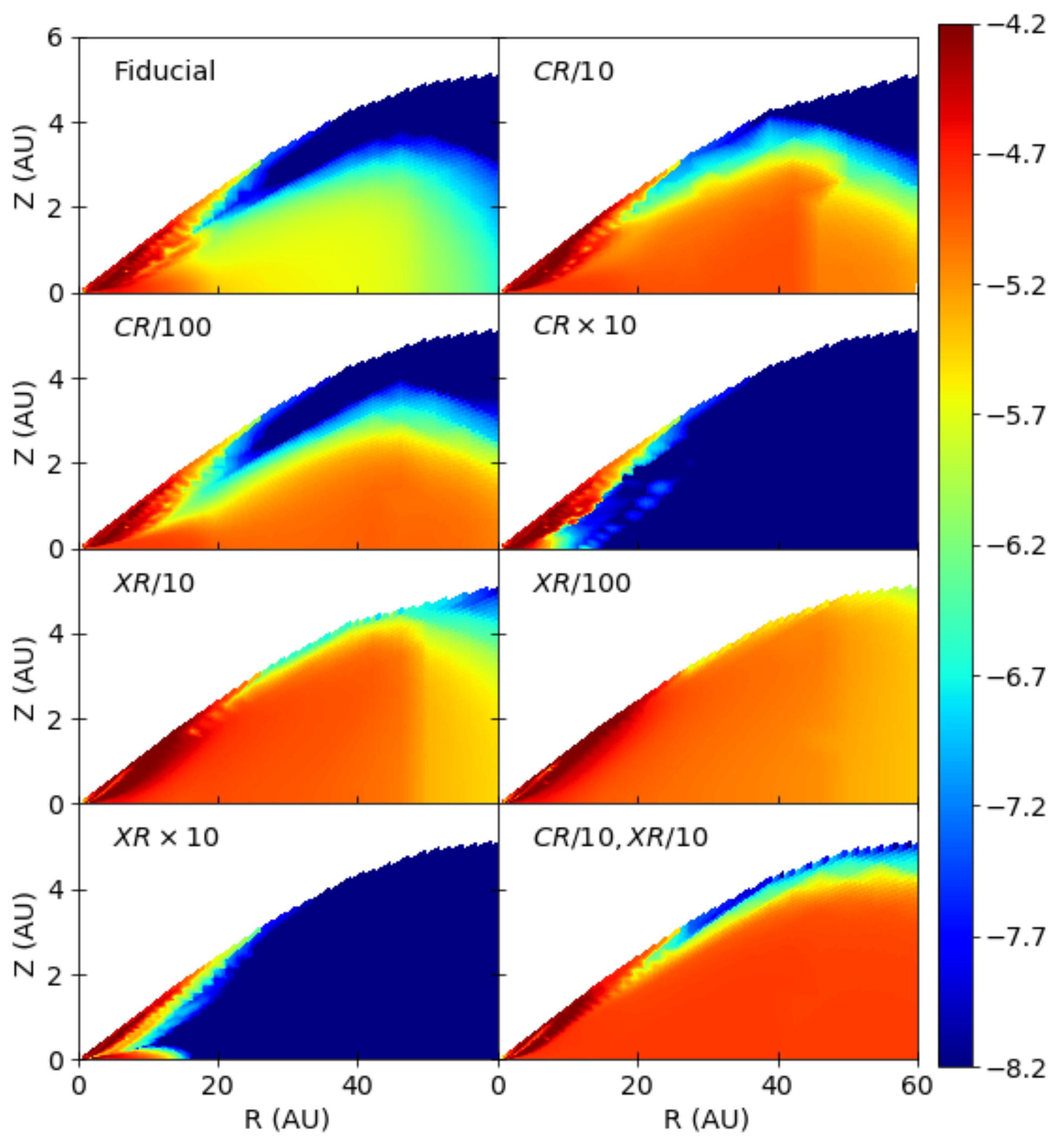}
\caption{Fractional CO abundances $f_{\rm CO}(R,z)$at 2~Myr. $f_{\rm CO} = N_{\rm CO} / \left(N_{\rm H} + 2 N_{\rm H_2} \right)$.}
\label{fig: COabun}
\end{figure}

Here we examine snapshots of our model disks after 2~Myr of disk
evolution, about the median age of stars in Taurus \citep{kraus09}. In
the fiducial model \citep[presented by][]{yu16}, CO depletion begins to
noticeably affect the outer disk ($R > 40$~au) by $\sim 1$~Myr, and
after 2~Myr the CO abundance has developed strong gradients in $R$ and
$z$. Figure \ref{fig: COabun} shows the CO fractional abundance
distribution $f_{\rm CO}(R,z)$, where $f_{\rm CO} = N_{\rm CO} /
\left(N_{\rm H} + 2 N_{\rm H_2} \right)$ ($f_{\rm CO}(R,z)
= 10^{-4}$ at t=0). Our results are qualitatively
similar to those of \citet{bosman18}, except for the exact
ionization-rate threshold at which CO destruction kicks in. While Bosman
et al.\ see no ionization-driven CO depletion for $\zeta_{\rm CR}
< 5 \times 10^{-18}$, we find midplane CO abundances that are at
least a factor of three lower than the ISM value of $f_{\rm CO} =
10^{-4}$ \citep{parvathi12} at $R > 15$~au in {\it all}
simulations---even in model CR/100, which has $\zeta_{\rm CR}
= 1.3 \times 10^{-19}$ at the disk surface. We also find factor-of-few
CO depletion beyond 15~AU in low-ionization models XR/100 and
CR/10,XR/10. The disappearance of CO in the outer disk seems to be a
robust conclusion.

Do any astrophysical disks receive lower X-ray or cosmic-ray fluxes than
represented in our simulations, possibly allowing CO gas to remain
intact? In the fiducial model, the X-ray ionization rates are set
according to the prescription of \citet{bai09} and are normalized using
an overall X-ray luminosity of $L_X = 2 \times 10^{30}$~erg~s$^{-1}$,
typical of a solar-mass star in the Taurus star-forming region
\citep{garmire00, telleschi07, robrade14}. The XR/100 model would
therefore represent a star with $L_X = 2 \times 10^{28}$~erg~s$^{-1}$.
\citet{telleschi07} find a positive correlation between X-ray luminosity
and stellar mass, so that the only stars in their sample with $L_X
\lesssim 2 \times 10^{28}$~erg~s$^{-1}$ have $M \lesssim 0.1
M_{\odot}$. We conclude that we are not overestimating the
X-ray fluxes of young solar-mass stars. 

For cosmic rays, \citet{cleeves13a} predict that the T-Tauri
magnetosphere strongly attenuates the incoming flux. They suggest that
the {\it maximum} reasonable value of $\zeta_{\rm CR}^{\rm H_2}$, the
ionization rate per H$_2$ molecule at the surface of a disk, in a
magnetic environment similar to the heliosphere is $1.4 \times
10^{-18}$~s$^{-1}$. In the ``T-Tauriosphere'' created by a typical
magnetically active young star, Cleeves et al.\ find $\zeta_{\rm
CR}^{\rm H_2} \la 10^{-20}$ at the disk surface.  Following
\citet{umebayashi09}, our fiducial disk has $\zeta_{\rm CR}^{\rm H_2} =
1.3 \times 10^{-17}$~s$^{-1}$ at the disk surface, which exponentially
decreases toward the midplane over a surface-density scale length of
96~g~cm$^{-2}$. In our CR/100 model, which has the lowest incident
cosmic-ray flux of any of our simulations, the cosmic ray-driven
ionization rate is still a factor of 10 above what Cleeves et al.\
recommend. It is possible that we are over-predicting the amount of
chemical depletion caused by cosmic rays.  However, since short-lived
radionuclides---which are not included separately in our
simulations---provide an ionization rate of $\zeta_{\rm RN} \sim (1-10)
\times 10^{-19}$~s$^{-1}$ at the disk midplane \citep{cleeves13b}, they
would become the dominant ionization source in the disk interior in our
CR/10, CR/100, and CR/10,XR/10 models. We thus conclude that the range
of midplane ionization rates we simulate is physically realistic, even
if the exact ionization source is unknown.

With a starting mass of only $0.015 M_{\odot}$ within 70~au of the
star---halfway between the minimum-mass solar nebula models of
\citet{weidenschilling77} and \citet{hayashi81}---the disk model
presented here is designed more for comparison with observations than
for realistic giant planet formation. While \citet{yu17a} found that the
combined effects of CO depletion in the outer disk, optically thick
emission in the inner disk, and the fact that the CO is not well modeled
by a single temperature combine to underestimate the mass of even this
low-mass disk, they also presented chemical models and synthetic
observations of a disk with $0.03 M_{\odot}$ within 70~au. The
higher-mass disk follows the same chemical evolution pathway as the disk
studied here, but with a time lag. After being irradiated by cosmic rays
and X-rays for $t$ and $t + 0.5$~Myr, respectively, the $0.015
M_{\odot}$ and $0.03 M_{\odot}$ disks have almost the same intensity
ratios of C$^{18}$O/$^{13}$CO J~=~2-1 and J~=~3-2, which are used as
mass diagnostics in the models of \citet{williams14} and
\citet{miotello16} (see Figure 8 of \citet{yu17a}). The effects of
ionization-driven chemistry are cumulative: given a long enough
radiation-exposure age, any disk should eventually lose most of its
gaseous CO.

\section{Radial re-distribution of CO}
\label{sec:distribution}

From Figure \ref{fig: COabun}, we see that CO depletion in the outer
disk is the most obvious consequence of continuous exposure to ionizing
radiation over million-year timescales. What is less obvious is that
after an initial $\sim 0.5$~Myr period of depletion, the CO
abundance near the disk surface in the inner 10~au recovers to reach its
initial value of $10^{-4}$ by $\sim 1.5$~Myr. Although the star's
dimming as it evolves along the Hayashi track cools the disk enough to
freeze most of the CO$_2$, there is a layer of CO$_2$ gas at the warm
disk surface at $R < 15$~au. This gaseous CO$_2$ layer provides the raw
material for the CO abundance rebound along the surface of the inner
disk:
\begin{equation}
{\rm He^+ + CO_2 \rightarrow CO^+ + O}
\label{eq:heplusco2}
\end{equation}
\begin{equation}
{\rm C^+ + CO_2 \rightarrow CO^+ + CO} 
\label{eq:cplusco2}
\end{equation}
\begin{equation}
{\rm H + CO^+ \rightarrow CO + H^+.}
\label{eq:hcoplus}
\end{equation}
Here we quantify the evolving radial distribution of the CO molecule.
We use the metric $M_{< 20} / M_{\geq 20}$, the ratio of the CO gas mass
contained inside $R = 20$~au to the CO gas mass located outside 20~au.
(Note that the disks simulated here have outer boundaries at $R =
70$~au.) As our disk begins its evolution, roughly equal portions of the
CO gas reside inside and outside the circle with radius 20~au. 

\begin{figure}[ht]
\centering
\includegraphics[width=0.9\textwidth]{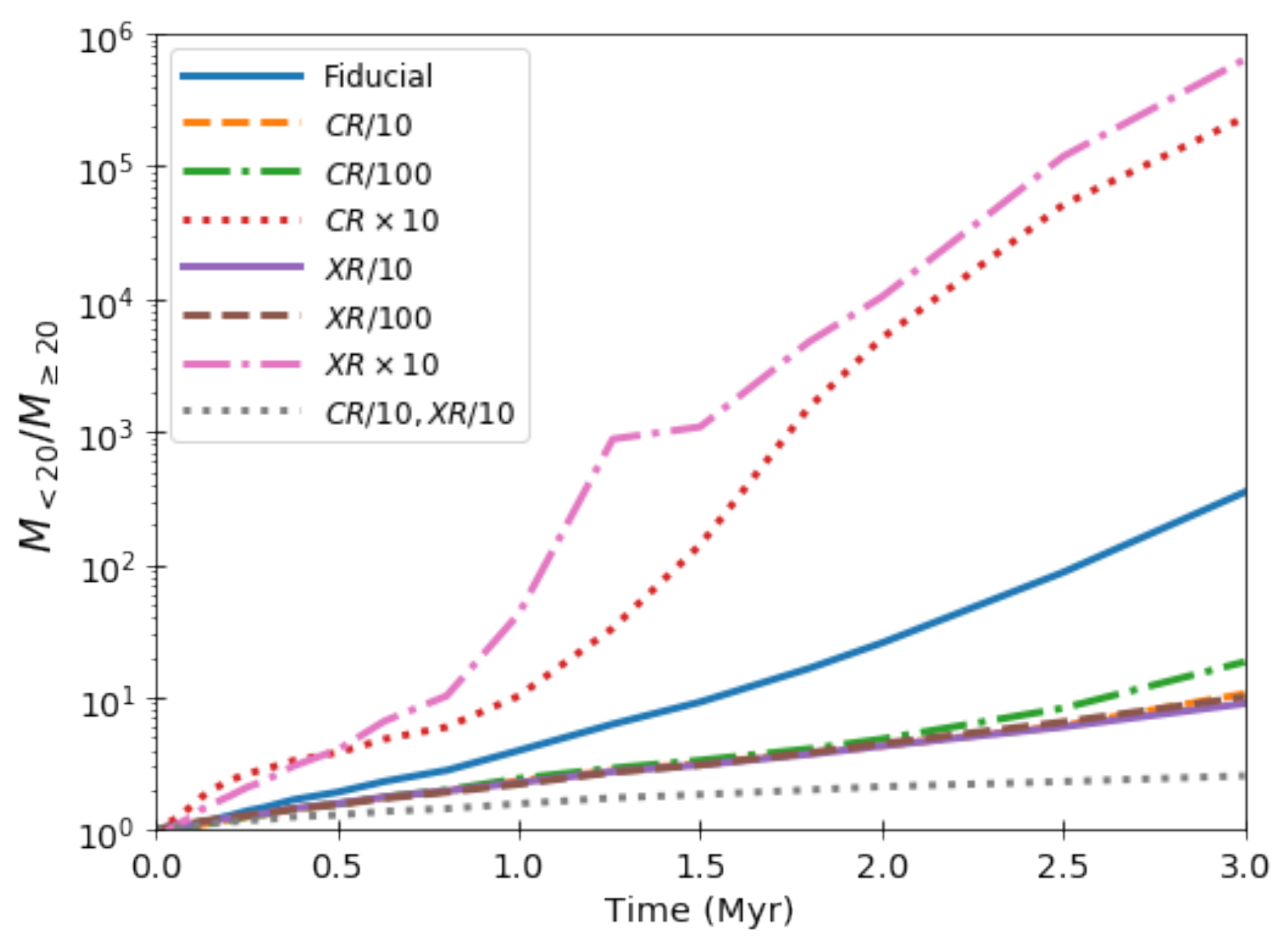}
\caption{$M_{< 20} / M_{\geq 20}$, the ratio of the CO mass inside 20~au
to the CO mass outside 20~au. Even in the models with the lowest
ionization rate, most of the CO is concentrated in the inner disk after
3~Myr of evolution. Disks with the highest ionization rates have the
most uneven radial distribution of CO.}
\label{fig:codist}
\end{figure}

Figure \ref{fig:codist} shows $M_{< 20} / M_{\geq 20}$ as a function of
time. Even in model CR/10,XR/10, which has the lowest ionization rate of
all our simulated disks, we still find that most of the CO resides
inside 20~au after several million years of evolution. The higher the
ionization rate (see Table \ref{ratestab}), the more the CO becomes
concentrated in the inner disk over time. 


Our results suggest that the CO emitting area could be an indirect
indicator of ionization rate: for disks of similar ages (i.e.\ those in
the same star-forming region), those with the most centrally peaked CO
emission (as measured in rare isotopologues such as C$^{18}$O and
C$^{17}$O) might see the highest flux of ionizing radiation. For
example, spatially resolved observations of the disk surrounding TW~Hya
show that the CO column density drops precipitously as a function of
radius, and that the drop occurs well inside the CO snow line
\citep{schwarz16}. The concentration of CO in the inner disk plus the
overall low CO abundance \citep[$f_{\rm CO} \leq 2.5 \times 10^{-6}$
everywhere in the disk;][]{bergin13, favre13, cleeves15, schwarz16,
huang18} suggest that CO may be succumbing to ionization-driven
depletion. To help distinguish between a CO-depleted gas disk and a
disk with interstellar CO/H$_2$ but small radial extent, we recommend
comparing the radial extents of the C$^{18}$O (J=3-2 or J=2-1) and dust
continuum emission. While many disks have $^{12}$C$^{16}$O haloes that
extend well into the surrounding molecular cloud, spatially resolved
observations of the CO-depleted disks IM~Lup and TW~Hya show that
C$^{18}$O has a similar extent to the dust disk
\citep[e.g.][]{cleeves16, schwarz16, huang18}. Disks that appear much
wider in C$^{18}$O than in millimeter dust continuum should not be
CO-depleted. For unresolved but inclined disks, wide line profiles can
indicate that the inner disk dominates the CO emission: in our fiducial
model, the velocity width of the CO $J=3-2$ line doubles over the course
of 3~Myr of evolution as the CO becomes more and more confined to the
inner disk \citep{yu17a}.



\section{Conclusions}
\label{sec:conclusions}

The conclusion from \citet{yu16} that chemical depletion can
substantially decrease the gas-phase CO abundance in the outer disk
after 1 Myr of evolution is robust against variations by factors of
1000 in the ionization rate from cosmic rays and X-rays. Our simulations
add to a growing body of literature predicting that CO abundances should
be sub-interstellar in T-Tauri disks \citep{aikawa97, bruderer12,
favre13, furuya14, reboussin15, drozdovskaya16, eistrup16, yu16,
eistrup18, schwarz18}, a finding that may explain why disk masses in
Lupus, $\sigma$~Orionis, and Chamaeleon I appear so low when measured
from $^{13}$CO and C$^{18}$O \citep{ansdell16, ansdell17, long17}.
Indeed, HD-based gas mass measurements are higher than CO-based
estimates for all disks with HD detections \citep{bergin13, mcclure16}.
In two well-resolved disks with observations in multiple CO
isotopologues---TW~Hya, discussed in \S \ref{sec:distribution}, and IM Lup
\citep{cleeves16}, comparison of thermo-chemical models with data
suggests that CO abundances are at least a factor of 20 below
interstellar.

To our prediction of CO chemical depletion beyond $\sim 20$~au, even in
disks with lower X-ray or cosmic-ray ionization rates than the
``standard'' values \citep[e.g.][]{garmire00, umebayashi09}, we add that
the radial distribution of CO may serve as an indirect tracer of
ionization rate. In our higher-ionization models (Fiducial,
CR$\times$10, XR$\times$10), gradual dissociation of CO$_2$ gas raises
the CO abundance in the inner 15~au of the disk, even as the molecule is
destroyed in the outer disk. The end result is a CO abundance
distribution that is sharply peaked in the inner disk.

This paper does not include any experiments with grain surface
properties, size distributions, or radial drift. While the reactions
that lead to CO dissociation happen in the gas phase, the reason that
gaseous CO does not re-form is because the constitutent atoms get locked
into hydrocarbons, which then freeze out on grain surfaces. Yet grain
growth, drift, and settling can lead to gaseous hydrocarbon ring
formation at the outer edge of the pebble disk, where the opacity is low
and UV photons can penetrate the full gas column \citep{bergin16}. Our
models should not be used to predict the behavior of either CO or
hydrocarbons in parts of the disk where the gas-to-solid ratio is more
than an order of magnitude different than the canonical value of 100
\citep{yu16}. Radial drift of cm-size pebbles followed by CO
desorption can also deplete the outer disk of CO gas
\citep[e.g.][]{krijt18}, an effect we have not explored. We also do not
investigate disks with dust traps, pressure maxima, or inner holes,
though it is becoming increasingly clear that few astrophysical disks
have smooth, power-law surface density distributions
\citep[e.g.][]{vandermarel13, casassus13, zhang14, vandermarel15, vandermarel16, canovas16}. 
While this work presents a useful framework
for analyzing the relationship between CO abundance and ionization rate,
models of astrophysical disks require require additional complexity to
accurately reproduce observations.


Acknowledgments: This research was performed in part at the Aspen Center
for Physics, which is supported by National Science Foundation grant
PHY-1607611. This research was partially funded by National Science
Foundation grant 1055910 to S.D.R. We acknowledge helpful conversations
with Nienke van der Marel and Richard Booth.


\clearpage


\end{document}